# Low-level cognitive skill transfer between two individuals' minds via computer game-based framework

Ahmet Orun
De Montfort University, Faculty of Computing, Engineering and media, Leicester UK
Email: aorun@dmu.ac.uk.  Phone: +44(0)116 3664408

**Abstract**

The novel technique introduced here aims to accomplish the first stage of transferring low-level cognitive skills between two individuals (e.g. from expert to learner) to ease the consecutive higher level declarative learning process for the target "learner" individual in a game environment. Such low-level cognitive skill is associated with the procedural knowledge and established at low-level of mind which can be unveiled and transferred by only a novel technique (rather than by a traditional educational environment ) like a highly interactive computer game domain in which a user exposes his/her unconscious mind behaviors via the game-hero non-deliberately during the game sessions. The cognitive data exposed by the game-hero would be recorded, and then be modelled by the artificial intelligence technique like Bayesian networks for an early stage of cognitive skill transfer and the cognitive stimuli are also generated to be used as game agents to train the learner.

## 1 Introduction

The cognitive model used within this work refers to  Adaptive Control of  Thought – Rational (ACT-R) a cognitive architecture  which was developed by Anderson at Carnegie Mellon University [1][2][3][4]. It suggests  that  two-level cognitive structure (low and high levels) of  the  human mind is effective for skill learning. The idea of ACT-R Cognitive architecture was also studied by Sun et al. [5]  which presents a skill learning model *Clarion*  specified as  a  bottom-up learning approach. According to their idea the procedural knowledge develops first  and declarative knowledge develops later.  The direct  cognitive skill development through the procedural level without declarative knowledge  inspires our suggested work on  low level cognitive  skill transfer  by using  a computer game domain.   Our technique uses a specifically selected computer game  in association with an artificial intelligence method called bayesian networks to identify the cognitive activity and  structural differences between  two individuals at  procedural level.

Several studies indicates that the procedural knowledge is more efficient to solve the problems than the declarative knowledge in the real-world domains.  In fact no any potential or type of declarative knowledge is efficient enough to model the real-world affairs accurately at very fundamental level. Hence the procedural knowledge/skill is inevitable to associate with all low level fundamental modelling process.  As far as the  main characteristics of  real-world domain is concerned;  it is unpredictable, mostly  consists of unique parameters and  difficult to model completely.

For this suggested work the selection of game is made  so that it includes as large number of attributes as possible to increase the discrimination power between the individuals. The game also should provide sufficient optional freedom  for the players to exhibits their personal cognitive characteristics in their behaviours [9]. The topic of game is not very important because during the play all content of  game scenario will already be converted to basic  materials  (e.g. cause-effect rules, symbols, emotional properties, etc.) to be located  at  procedural level. This stage links to cognitive learning process.  This level  also covers  an area which is free from any  declarative type of  topics, knowledge or actions (this structure is at some extent  similar to configuration of a computer  where there is binary machine language at the bottom and words, pictures, etc. at the top level).   The symbols at the procedural level are effectively associated with the subconscious mind and may be represented by any  object metaphorically  (e.g. tiger may correspond to passion, etc.) which may be represented  in unconscious mind  by these forms [11].

## 2 Suggested model

The learning and action stages are shown by the separate configurations in Figure 1 with reference to ACT-R cognitive architecture. For the learning stage the suggested model is based on an assumption that the real-world data are received and interpreted by human mind by two-fold categorization as; 1. declarative knowledge (merged with substantial objects, actions or concepts) and 2. procedural knowledge (refer to unexplainable subconscious materials). In this work we rather focus on the high level complexity of procedural level of mind and its formation during the game activities. Even though the action stage seems like to be highly related to declarative level, it certainly provides some clues about the cognitive skills which belong to procedural level. But this invisible connections can only be exposed by an artificial intelligence techniques like bayesian or neural networks. The use of Bayesian network in modelling cognitive processes was also introduced by Lee [6] where four of the most important potential hierarchical Bayesian contributions to cognitive modelling were discussed. Huan [7] also conducts a research simulating the visual cognitive behaviour of a designer by using the neural networks. Once the cognitive skills located at procedural level are identified, they are transferred to another individual by learning action through the game domain (Figure 2).

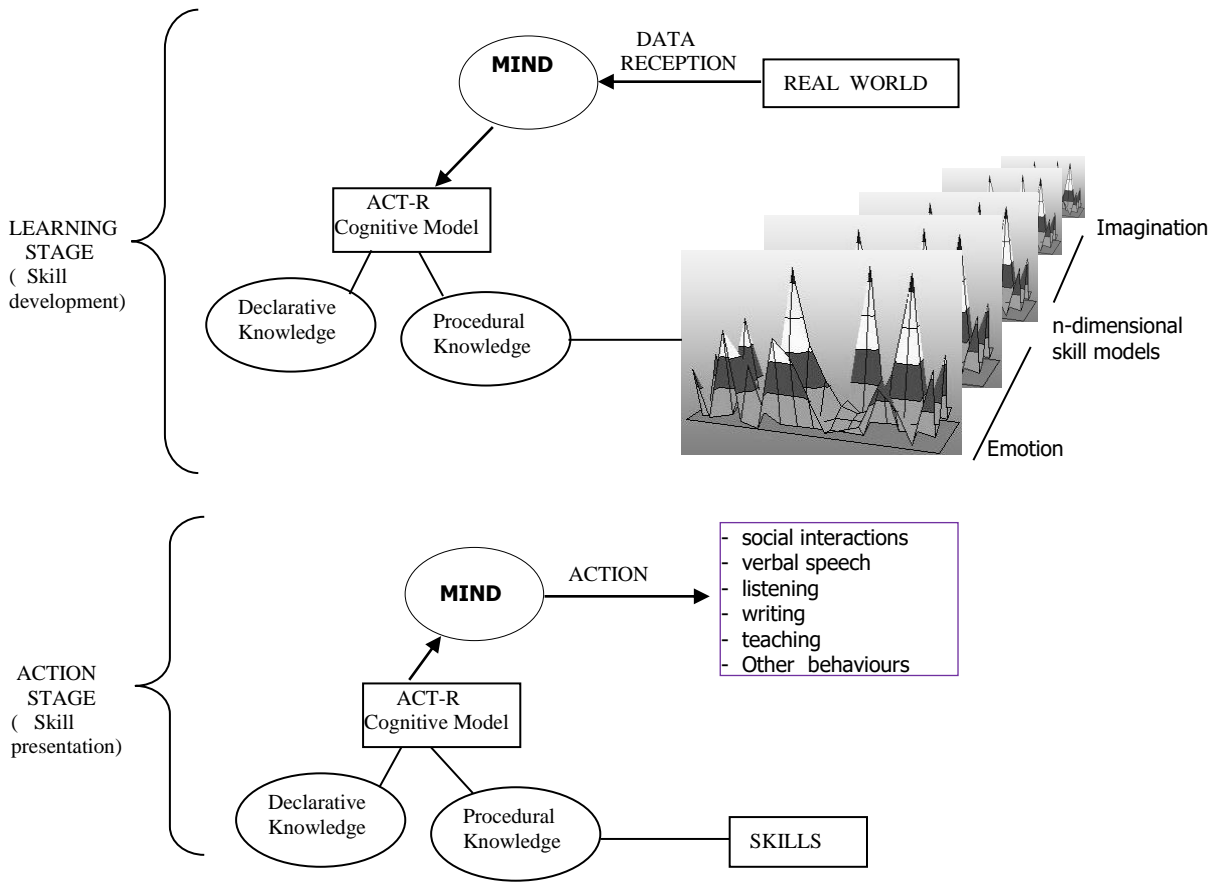

Fig. 1. Simplified cognitive functions of mind during the skill development and skill presentation stages with reference to ACT-R Cognitive Model.

## 3 The use of a game domain

Several studies suggest that using a computer game domain may help unveil cognitive characteristics of human mind for personal identification or psychotherapeutic treatment. In example Robertson and Miller [8] indicate the significant role of game consoles in learning skill development of children. Orun and Seker [9] also conduct a study to distinguish between two individuals by their cognitive behaviour characteristics exhibited in the game domain. Bostan [10] attempts to predict goal-directed behaviour of both player and non-player characters in a computer game for discovering the opportunities of using a Player and Agent Personality Database (PAPD. An indirect interaction between two individuals via game domain and training domain is shown in Figure 2 with regards to two-layer cognitive structure. Our suggested model rather focus on 1. phase specified as low-level cognitive skill transfer as a part of whole skill transfer process between the individuals.

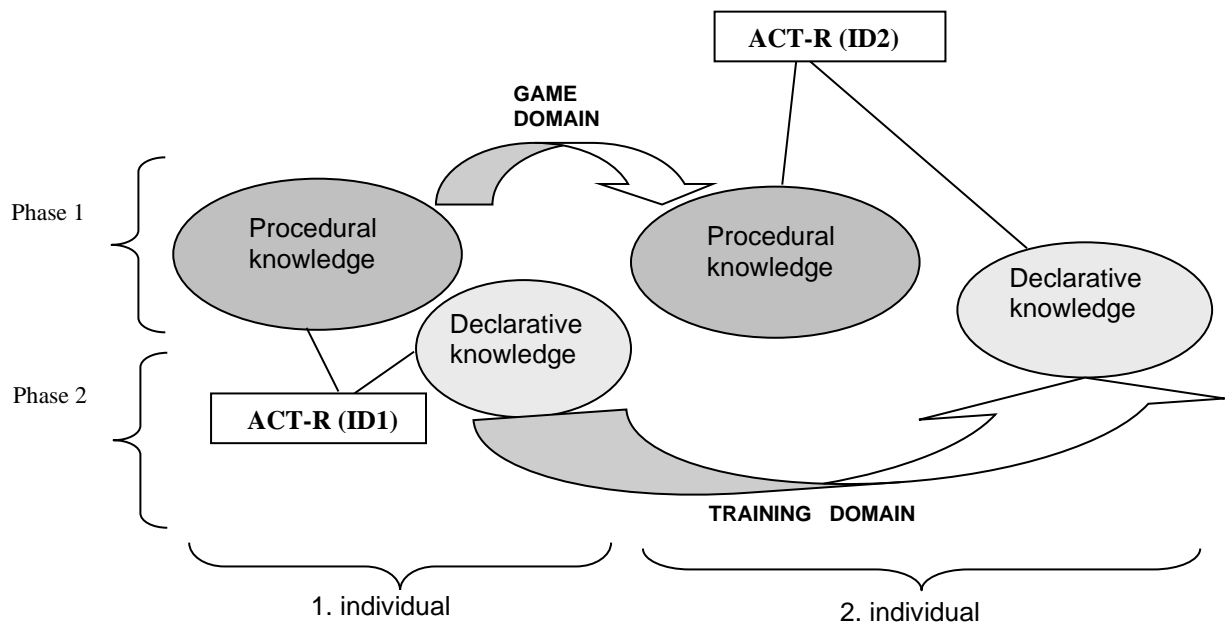

Fig. 2. The suggested model help transfer low-level cognitive knowledge between two individuals by using a game domain which later may ease high-level learning process of declarative knowledge between the same individuals.

The operational diagram of suggested system is shown in Figure 3 in which Cognitive skill transfer process loop continues by using game stimuli (game agents) until the difference between two individual reaches to minimum level where no any detection of distinction between them can be made by bayesian classifier. Basically the bayesian inference tool detects the discriminant attributes and configure their links by a network. Then bayesian classifier make an assessment of similarity between the two classes (here two individuals) over the attributes predefined. By the process shown in Figure 3, it is assumed that the more similarity between two individuals the higher level of cognitive skill transfer achieved.

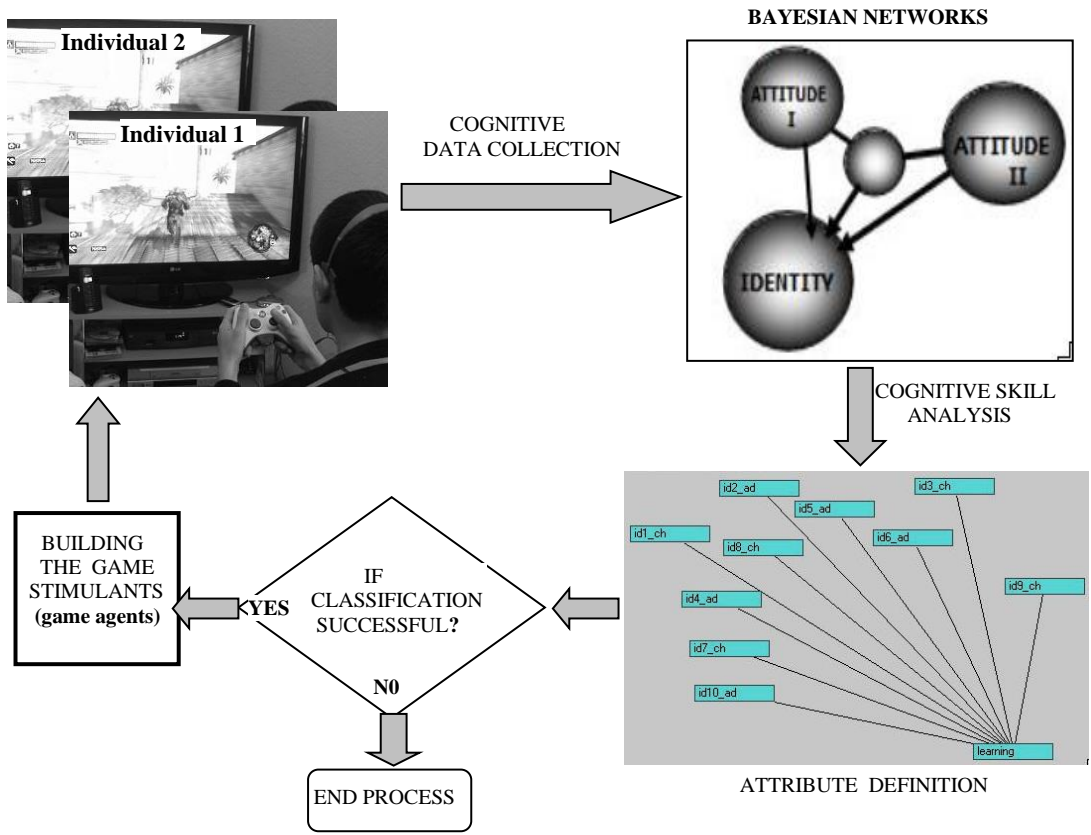

Fig. 3. Cognitive skill transfer process continues by using game stimuli (game agents) until the difference between two individual reaches to minimum level where no any detection of distinction between them can be made by bayesian classifier.

## 4 Results and discussion

### 4.1 Identification stage

The data set is built by naked-eye video interpretation of the game scenario played by player 1 and 2. Each behavioural action in the game (e.g. fighting, horse riding, climbing, etc.) is labelled as an attribute to form data set. Then whole data set was divided by two for training/test purposes in Bayesian network utility (PowerPredictor™). This highly efficient classification utility help expose the differences between the cognitive behavioural characteristics of both players as shown in Table 1 and Figure 4. The classification procedure also yields the network configuration (Figure 4) with classification accuracy of 87.5%. It means that the discrimination of the game users by the network is accomplished by the attributes automatically selected by the classifier utility. The classification accuracy further proves robustness of the data acquisition and analysis method and reliability of the model and results obtained. The behavioural difference are identified by the attributes below as labelled by behaviours' index numbers between 1 and 10.

- **(1)** Fighting      [ refers to fighting action in the game domain ]
- **(2)** obstacle      [ refers to attitude of player to tackle with a physical obstacle ]
- **(3)** riding_hrs    [ refers to horse riding ]
- **(4)** facing_sol    [refers to attitude of player to face a soldier in the game ]
- **(5)** climbing      [ any climbing action (e.g. tower, building, hill, etc.)
- **(6)** location      [player location ]
- **(7)** facing_prs    [facing a person ]
- **(8)** movement      [walking, running ]
- **(9)** listening     [looking at face, in motion]
- **(10)** attack_civ   [attacking civilians ]

Table 1 also shows how any same type of condition or stimuli in the game domain effects the player 1 and 2. This information of table content is extracted from the networks combinations built after the chain of experiments using Bayesian inference tool (PowerConstructor$^{TM}$). The numbers in parantheses refer to index numbers of players' behaviours.

Table I - Behavioural differences between the game players (Superscript numbers on each row indicate the codes of the Bayesian Network nodes' abbreviations)

| EFFECT OF CONDITIONS/ STIMULIS IN GAME ENVIRONMENT | ID 1 (PLAYER 1) | ID 2 (PLAYER 2) | BAYESIAN NET NODE ABBREVIATION |
|---|---|---|---|
| Location (indoor)$^6$ of player | Does not affect any of behaviours | Leads player to Walk$^8$ indoor, run$^8$ outdoor | Location$^6$ Movement$^8$ |
| Looking at face$^9$ | Causes attacking the soldiers$^4$ | Causes more than one behaviours: horse riding, climbing, attacking civilians | Listen-2$^9$ |
| Climbing$^5$ | No any social behaviour connection | Causes attacking civilians$^{10}$ | Climbing$^5$ facing_prs$^7$ |
| Facing a physical obstacle$^2$ | Connected to fighting actions$^1$ | Connected to verbal communication style$^9$ (face-to–face) | Obstacle$^2$ Fighting$^1$ listen_2$^9$ |
| Horse riding$^3$ | Causes attacking soldiers$^4$ | Connected to verbal communication style$^9$ (face-to–face) | riding_hrs$^3$ facing_sol$^4$ |

In Figure 4 the class node (ID) refers to player 1 and 2 to be identified by the classification process. The nodes left unconnected are not included into the network this means that their inclusion do not make any further contribution to improve classification accuracy. This initial result of high classification accuracy (87.5%) refers to early stage of skill transfer procedure (identification stage) where system can distinguish between the two individuals via those five attributes (e.g. fighting, obstacle, etc.).

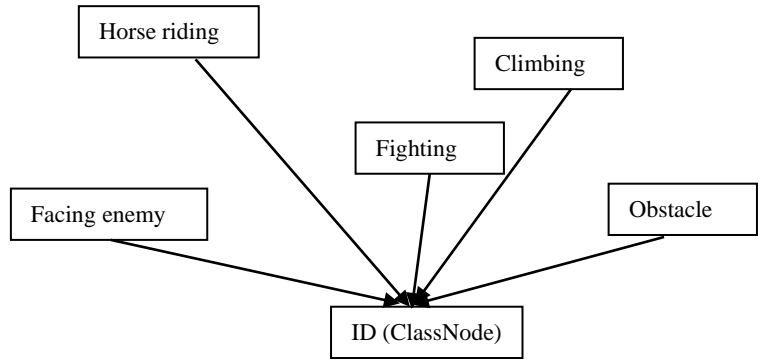

Figure 4. The Bayesian Network that yielded the highest accuracy 87.5% and represents the interaction between the attributes identified for both players' cognitive behaviours which are explained in Table 1.

## 4.2 Cognitive skill transfer stage

At this stage the stimulant game agents will be designed and embedded into a graphic game domain on the basis of those five attributes for cognitive skill transfer. For example for attribute "*fighting*" the stimulant game agent will act to encourage the second player (learner) to behave exactly like first player (expert) by an award-penalty game process. In Figure 6 the behavioural curves of both players effected by stimuli of game conditions / actions in relation with Table 2. Player 1 and 2 cited as ID1 and ID2 respectively. The graph of curves is presented to show how the behaviours of two individuals may be closer to each other gradually by the effects of game stimulant. But we have to note that at this stage no any particular game agent is used yet as a stimulant but the graph shows the possible interaction between the user behaviours and the game conditions/actions which may be used as stimulants. As the proposed technique aims to make second individual (learner) to gain cognitive skill from the first individual (expert), the desired behavioural curves will be similar to those shown in Figure 6. The only difference will be that the horizontal axis should refer to "time sequence" instead of "stimuli" effects which corresponds to skill transfer process loop shown in Figure 3.

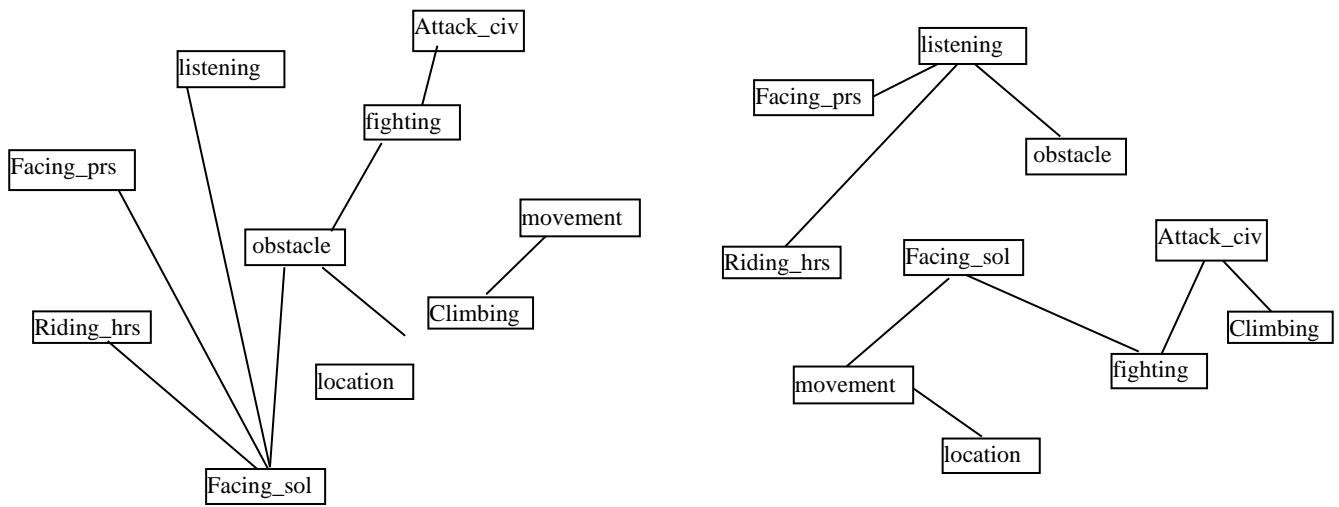

Fig.5. Bayesian Networks showing the connection between the attributes identified for the cognitive behaviours which belong to Player 1 (on the left) and Player 2 (on the right).

The information of behavioural links for both players was extracted from the networks as shown in Fig. 5. The attributes refer the attribute's index number shown in Table 1. The behaviours of players (shown as ID1 and ID2) correspond to players options as given in Table 1. Behavioural curves of both players are also shown in Fig.6

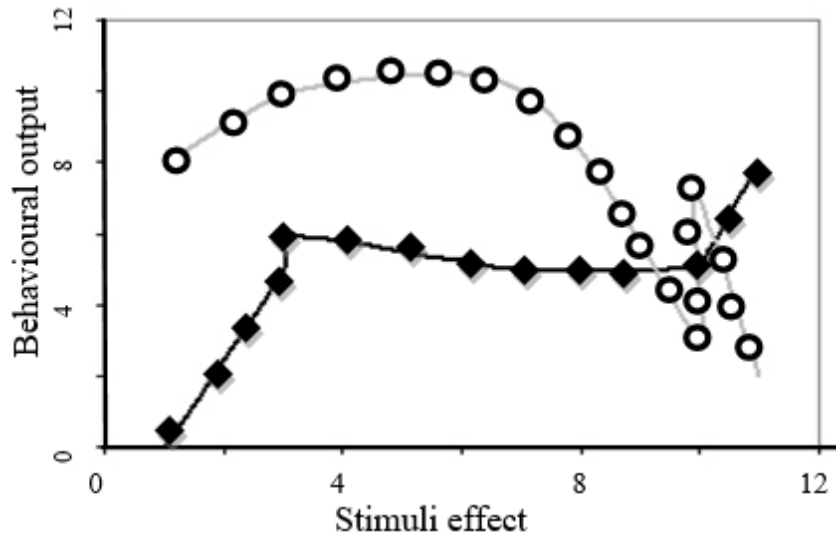

Fig. 6. Behavioural output curves for Player 1 (diamond) and Player 2 (circle) against the stimuli effects of the game actions and conditions which are linked to Table 1.

## Conclusion

Within this work the proposed technique demonstrated that low-level cognitive skills tranfer between two individuals (e.g. sudent and lecturer, expert and learner, etc.) would be possible at procedural level to ease the consecutive higher level declarative knowledge transfer between the same individuals in a specific training environment. Such low-level cognitive skill is associated with the procedural (*intrinsic*) knowledge and located at low-level of mind which can be unveiled and transferred by only a non-conventional technique other than a traditional educational environment. This may be done by a highly interactive computer game domain in which a user exposes his/her unconscious mind by the cognitive behaviours undeliberately during the gane session. These cognitive data can then be modelled by an artificial intelligence technique like Bayesian networks, Neural Nets, etc. for an early stage of following cognitive skill transfer.